\magnification \magstep1
\raggedbottom 
\openup 2\jot
\voffset6truemm
\def\II{{\rm1\!\hskip-1pt I}}
\centerline {\bf ELLIPTICITY CONDITIONS FOR THE LAX OPERATOR}
\centerline {\bf OF THE KP EQUATIONS}
\vskip 1cm
\leftline {Giampiero Esposito$^{1,2}$ and Boris G. Konopelchenko$^{3,4}$}
\vskip 0.3cm
\noindent
{\it ${ }^{1}$Istituto Nazionale di Fisica Nucleare, Sezione
di Napoli, Complesso Universitario di Monte S. Angelo,
Via Cintia, Edificio N', 80126 Napoli, Italy}
\vskip 0.3cm
\noindent
{\it ${ }^{2}$Universit\`a di Napoli Federico II, Dipartimento
di Scienze Fisiche, Complesso Universitario di Monte S. Angelo,
Via Cintia, Edificio N', 80126 Napoli, Italy}
\vskip 0.3cm
\noindent
{\it ${ }^{3}$Dipartimento di Fisica, Universit\`a di Lecce,
Via Arnesano, 73100 Lecce, Italy}
\vskip 0.3cm
\noindent
{\it ${ }^{4}$Istituto Nazionale di Fisica Nucleare, Sezione di
Lecce, 73100, Italy}
\vskip 1cm
\noindent
{\bf Abstract.} The Lax pseudo-differential operator plays a key
role in studying the general set of KP equations, although it is
normally treated in a formal way, without worrying about a complete
characterization of its mathematical properties. The aim of the
present paper is therefore to investigate the ellipticity condition.
For this purpose, after a careful evaluation of the kernel with the
associated symbol, the majorization ensuring ellipticity is 
studied in detail. This leads to non-trivial restrictions on the
admissible set of potentials in the Lax operator. When their time
evolution is also considered, the ellipticity conditions turn out
to involve derivatives of the logarithm of the $\tau$-function. 
\vskip 100cm
\leftline {\bf 1. Introduction}
\vskip 0.3cm
\noindent
Several important developments in modern mathematical physics are
due to the investigation of pseudo-differential operators on 
${\bf R}^{m}$ and on general Riemannian manifolds [1]. For our
purposes, it is sufficient to recall the following basic properties.
\vskip 0.3cm
\noindent
(i) A linear partial differential operator $P$ of order $d$ can be
written in the form
$$
P \equiv \sum_{|\alpha| \leq d} a_{\alpha}(x)D_{x}^{\alpha}
\eqno (1.1)
$$
where (here $i \equiv \sqrt{-1}$)
$$
|\alpha| \equiv \sum_{k=1}^{m}\alpha_{k}
\eqno (1.2)
$$
$$
D_{x}^{\alpha} \equiv (-i)^{|\alpha|}
{\left({\partial \over \partial x_{1}}\right)}^{\alpha_{1}}
...
{\left({\partial \over \partial x_{m}}\right)}^{\alpha_{m}}
\eqno (1.3)
$$
and $a_{\alpha}$ is a $C^{\infty}$ function on ${\bf R}^{m}$ for
all $\alpha$. The associated {\it symbol} is, by definition,
$$
p(x,\xi) \equiv \sum_{|\alpha| \leq d}a_{\alpha}(x)\xi^{\alpha}
\eqno (1.4)
$$
i.e. it is obtained by replacing the differential operator 
$D_{x}^{\alpha}$ by the monomial $\xi^{\alpha}$. The pair 
$(x,\xi)$ may be viewed as defining a point of the cotangent 
bundle of ${\bf R}^{m}$, and the action of $P$ on the elements 
of the Schwarz space $\cal S$ of smooth complex-valued 
functions on ${\bf R}^{m}$ of rapid decrease is given by
$$
Pf(x) \equiv \int {\rm e}^{i(x-y)\cdot \xi}p(x,\xi)f(y)dy d\xi
\eqno (1.5)
$$
where the $dy=dy_{1}...dy_{m}$ and $d\xi=d\xi_{1}...d\xi_{m}$ orders
of integration cannot be interchanged, since the integral is not
absolutely convergent.
\vskip 0.3cm
\noindent
(ii) Pseudo-differential operators are instead a more general class
of operators whose symbol need not be a polynomial but has suitable 
regularity properties. More precisely, let $S^{d}$ be the set of
all symbols $p(x,\xi)$ such that [1]
\vskip 0.3cm
\noindent
(1) $p$ is smooth in $(x,\xi)$, with compact $x$ support.
\vskip 0.3cm
\noindent
(2) For all $(\alpha,\beta)$, there exist constants 
$C_{\alpha,\beta}$ for which
$$
\left | D_{x}^{\alpha}D_{\xi}^{\beta}p(x,\xi) \right |
\leq C_{\alpha,\beta} (1+|\xi|)^{d-|\beta|}
\eqno (1.6)
$$
for some {\it real} (not necessarily positive) 
value of $d$, where $|\beta| \equiv
\sum_{k=1}^{m} \beta_{k}$ (see (1.2)). The associated 
{\it pseudo-differential operator}, defined on the Schwarz space
and taking values in the set of smooth functions on ${\bf R}^{m}$
with compact support:
$$
P: {\cal S} \rightarrow C_{c}^{\infty}({\bf R}^{m})
$$
is defined in a way formally analogous to Eq. (1.5).
\vskip 0.3cm
\noindent
(iii) Let now $U$ be an open subset with compact closure in
${\bf R}^{m}$, and consider an open subset $U_{1}$ whose closure
${\overline U}_{1}$ is properly included into $U$: 
${\overline U}_{1} \subset U$. If $p$ is a symbol of order $d$
on $U$, it is said to be {\it elliptic} on $U_{1}$ if there exists
an open set $U_{2}$ which contains ${\overline U}_{1}$ and
positive constants $C_{i}$ so that
$$
|p(x,\xi)|^{-1} \leq C_{1} (1+|\xi|)^{-d}
\eqno (1.7)
$$
for $|\xi| \geq C_{0}$ and $x \in U_{2}$, where
$$
|\xi| \equiv \sqrt{g^{ab}(x)\xi_{a}\xi_{b}}
=\sqrt{\sum_{k=1}^{m} \xi_{k}^{2}}.
\eqno (1.8)
$$
The corresponding operator $P$ is then elliptic.
\vskip 0.3cm
{}From a mathematical point of view, pseudo-differential operators
occur in many problems in global analysis [1, 2],
and recent developments deal with the functional calculus of
pseudo-differential boundary-value problems [3]. From a physical
point of view, such a formalism is important in quantum gravity and
quantum field theory [4--6]. In particular, we are here interested
in an interdisciplinary field, i.e. a rigorous approach to the
Kadomtsev--Petviashvili (hereafter KP) equations. 
Recall that the KP equation can be written in the form [7, 8]
$$
{\partial \over \partial x} \left({\partial u \over
\partial t}+6u {\partial u \over \partial x}
+{\partial^{3}u \over \partial x^{3}}\right)
=3 \alpha^{2}{\partial^{2}u \over \partial y^{2}}.
\eqno (1.9)
$$
If $\alpha^{2}=1$, it describes an Hamiltonian wave system which
is exactly solvable but not Liouville integrable. It exhibits a
degenerative dispersion, the asymptotic states for $t \rightarrow
\pm \infty$ do not coincide and an infinite number of invariants
of motion exist. If $\alpha^{2}=-1$, it describes an Hamiltonian
wave system exactly solvable and completely integrable. It exhibits
non-degenerative dispersion and lack of decay, and the asymptotic
states for $t \rightarrow \pm \infty$ coincide upon imposing 
rapid-decrease boundary conditions. Moreover, the number of invariants
of motion remains infinite [8].

The general set of KP equations may be described by using
the first-order operator
$$
T \equiv {\partial \over \partial x}=\partial_{x}
\eqno (1.10)
$$
with $x \in {\bf R}$, and the associated Lax pseudo-differential
operator
$$
L \equiv T+\sum_{k=1}^{\infty}u_{k}(x,t_{1},t_{2},t_{3},...)T^{-k}
\eqno (1.11)
$$
where the functions $u_{k}$ are here called the `potentials'.
By doing so, one allows in general for their dependence on an
infinite number of time variables $t \equiv (t_{1},t_{2},...,
t_{p},...)$. 
On assuming that $T^{-1}$ is a well defined inverse operator (see
section 2), so that
$$
T \cdot T^{-1}=T^{-1} \cdot T= \II
\eqno (1.12)
$$
one can compose the Lax operator with itself, giving rise to its
`powers', i.e. $L^{n} \equiv L \cdot L^{n-1}$, for all
$n=2,3,...,\infty$. Each such power has a differential part, denoted
by $B_{n}$. To begin one sets
$$
B_{1} \equiv T
\eqno (1.13)
$$
and, by virtue of (1.12), one finds
$$
B_{2} \equiv T^{2}+2u_{1}
\eqno (1.14)
$$
$$
B_{3} \equiv T^{3}+3u_{1}T+3(u_{2}+u_{1,x})
\eqno (1.15)
$$
and so on. The KP hierarchy of integrable equations is then 
defined by the generalized Lax equation [9, 10] 
$$
{\partial L \over \partial t_{n}}=[B_{n},L]=B_{n}L-LB_{n}
\eqno (1.16)
$$
and by the Zakharov--Shabat equation
$$
{\partial B_{m}\over \partial t_{n}}
-{\partial B_{n}\over \partial t_{m}}=[B_{n},B_{m}]
\eqno (1.17)
$$
which may be seen as the compatibility conditions of the linear
equations
$$
L\psi=\lambda \psi
\eqno (1.18)
$$
and
$$
{\partial \psi \over \partial t_{n}}=B_{n}\psi
\eqno (1.19)
$$
for all $n$, under the assumption that
$$
{\partial \lambda \over \partial t_{n}}=0.
\eqno (1.20)
$$
At this stage, since $t_{1}$ plays the same role as $x$, $t_{1}$
or $x$ are used without distinction in the literature [10].
Once the equations (1.16) are written for all values of $n$, the
coefficients of $T^{-k}$ are equated, and this leads to an infinite
set of equations
$$
{\partial u_{k}\over \partial t_{n}}
=\varphi_{kn}
\eqno (1.21)
$$
where $\varphi_{kn}$ are certain differential polynomials in the
potentials and their derivatives. For example, from the equations [10]
$$
{\partial u_{1}\over \partial t_{2}}=u_{1,xx}+2u_{2,x}
\eqno (1.22)
$$
$$
{\partial u_{2}\over \partial t_{2}}=u_{2,xx}+2u_{3,x}
+2u_{1}u_{1,x}
\eqno (1.23)
$$
$$
{\partial u_{1}\over \partial t_{3}}=u_{1,xxx}+3u_{2,xx}
+3u_{3,x}+6u_{1}u_{1,x}
\eqno (1.24)
$$
one gets the KP equation for $u_{1}$:
$$
{\partial \over \partial x}\left(4{\partial u_{1}\over \partial t_{3}}
-12u_{1}{\partial u_{1}\over \partial x}
-{\partial^{3}u_{1}\over \partial x^{3}}\right)
-3{\partial^{2}u_{1}\over \partial t_{2}^{2}}=0.
\eqno (1.25)
$$

In section 2 we derive the kernel of the Lax pseudo-differential operator
by using a Green-function method and the theory of distributions
when all potentials only depend on $x$.
In section 3 we obtain the symbol from the kernel of section 2, and 
derive suitable majorizations which ensure ellipticity of the
Lax operator. Strong ellipticity is studied in section 4. 
Behaviour of the ellipticity conditions under KP flows is investigated
in section 5, and concluding remarks are presented in section 6,
while the appendix describes relevant details.
\vskip 0.3cm
\leftline {\bf 2. The Lax operator and its kernel}
\vskip 0.3cm
\noindent
Following [10], we first consider a `restricted' form of the Lax
operator, for which the potentials $u_{k}$ only depend on $x$. The
general form (1.11) will be restored in section 5, where time
evolution is studied (cf section 3 of [10]).

Once the operator (1.10) is given, the inverse operator $T^{-1}$ is an
integral operator with kernel given by the Green function 
$G_{1}(x,y)$ of $T$. Its action on any function $f$ in its domain reads
$$
(T^{-1}f)(x)=\int_{-\infty}^{\infty} G_{1}(x,y)f(y)dy
\eqno (2.1)
$$
where the Green function $G_{1}$ obeys the equation
$$
T_{x}G_{1}(x,y)=\delta(x,y).
\eqno (2.2)
$$
More precisely, the Green function $G_{1}$ is a kernel which 
solves the equation
$$
{\partial \over \partial x}G_{1}(x,y)=0 \; \; \forall 
\; \; x \not = y
\eqno (2.3)
$$
and the jump condition [11]
$$
\lim_{x \to y^{+}}G_{1}(x,y)-\lim_{x \to y^{-}}G_{1}(x,y)=1.
\eqno (2.4)
$$
The problem described by Eqs. (2.3) and (2.4) is solved by
$$
G_{1}(x,y)=A_{1,1}(y) \; \; {\rm if} \; \; x > y
\eqno (2.5a)
$$
$$
G_{1}(x,y)=A_{1,1}(y)-1 \; \; {\rm if} \; \; x < y
\eqno (2.5b)
$$
where $A_{1,1}(y)$ is an arbitrary smooth function of $y$ unless a
suitable boundary condition is specified (see below).

Similarly, the operators $T^{-2}, T^{-3}$ and so on are integral
operators with kernel given by the Green function of 
$T^{2},T^{3},...,$ respectively. For example, the operator 
$T^{2}=T \; T$ has a Green function $G_{2}(x,y)$ satisfying the
differential equation
$$
{\partial^{2}\over \partial x^{2}}G_{2}(x,y)=0 \; \; 
\forall \; \; x \not = y
\eqno (2.6)
$$
the continuity condition 
$$
\lim_{x \to y^{+}}G_{2}(x,y)=\lim_{x \to y^{-}}G_{2}(x,y)
\eqno (2.7)
$$
and the jump condition
$$
\lim_{x \to y^{+}}{\partial G_{2}\over \partial x}
-\lim_{x \to y^{-}} {\partial G_{2}\over \partial x}=1.
\eqno (2.8)
$$
Equations (2.6)--(2.8) are solved by
$$
G_{2}(x,y)=A_{1,2}(y)+A_{2,2}(y)x \; \; {\rm if} 
\; \; x > y
\eqno (2.9a)
$$
$$
G_{2}(x,y)=y+A_{1,2}(y)+(A_{2,2}(y)-1)x \; \; {\rm if}
\; \; x < y
\eqno (2.9b)
$$
where now two arbitrary functions $A_{1,2}$ and $A_{2,2}$ are
involved because $G_{2}$ is the Green function of a second-order
differential operator. 

It is therefore clear that, assuming for the time being that $u_{k}$
only depends on $x$, the Lax operator
(1.11) can be viewed as an integral operator whose action 
is given by 
$$
(L \psi)(x)=\int_{-\infty}^{\infty}K(x,y)\psi(y)dy
\eqno (2.10)
$$
with kernel
$$
K(x,y)=-\delta'(x,y)+\sum_{p=1}^{\infty}u_{p}(x)G_{p}(x,y)
\eqno (2.11)
$$
where we have used the well known distributional action of the first
derivative of the Dirac delta functional [12], and the Green function
$G_{p}(x,y)$ can be expressed in the form
$$
G_{p}(x,y)=\sum_{r=1}^{p}C_{r,p}(y)x^{r-1}.
\eqno (2.12)
$$
The `coefficients' $C_{r,p}$ are actually functions of $y$ obeying
a law of the type (see (2.5) and (2.9))
$$
C_{r,p}(y)=A_{r,p}(y) \; \; {\rm if} \; \; x>y
\eqno (2.13a)
$$
$$
C_{r,p}(y)=B_{r,p}(y) \; \; {\rm if} \; \; x<y
\eqno (2.13b)
$$
where the coefficients $B_{r,p}(y)$ in (2.13b) can be expressed in
terms of the coefficients $A_{r,p}(y)$ after imposing the
continuity conditions
$$
\lim_{x \to y^{+}}{\partial^{q}G_{p}\over \partial x^{q}}
-\lim_{x \to y^{-}}{\partial^{q}G_{p}\over \partial x^{q}}=0
\; \; \forall \; \; q=0,1,...,p-2
\eqno (2.14)
$$
and the jump condition
$$
\lim_{x \to y^{+}}{\partial^{p-1}G_{p}\over \partial x^{p-1}}
-\lim_{x \to y^{-}}{\partial^{p-1}G_{p}\over \partial x^{p-1}}
=1 .
\eqno (2.15)
$$

In [7], the Green function $G_{1}(x,y)$ given by (2.5a) and (2.5b)
has been written in the form
$$
G_{1}(x,y)={1\over 2}\Bigr[\theta(x-y)-\theta(y-x)\Bigr]
\eqno (2.16)
$$
where $\theta$ is the step function such that $\theta(x)=1$ if
$x>0$, $\theta(0)={1\over 2}$, $\theta(x)=0$ if $x<0$. Equation (2.16)
corresponds to choosing 
$$
A_{1,1}(y)={1\over 2}
\eqno (2.17)
$$
in Eqs. (2.5a) and (2.5b). The operator $T^{-1}$ is then the integral
operator
$$
T^{-1}:f \rightarrow {1\over 2} \int_{-\infty}^{x}f(y)dy
-{1\over 2}\int_{x}^{\infty}f(y)dy.
\eqno (2.18)
$$
Similarly, the operator $T^{-2}$ turns out to be the integral operator
$$ \eqalignno{
\; & T^{-2}: f \rightarrow {1\over 4}\int_{-\infty}^{x}dy
\int_{-\infty}^{y}f(z)dz
-{1\over 4}\int_{-\infty}^{x}dy
\int_{y}^{\infty}f(z)dz \cr
&-{1\over 4}\int_{x}^{\infty}dy \int_{-\infty}^{y}f(z)dz
+{1\over 4}\int_{x}^{\infty}dy \int_{y}^{\infty}f(z)dz.
&(2.19)\cr}
$$
By comparison with (2.9a) and (2.9b) this leads to the evaluation of
$A_{1,2}(y)$ and $A_{2,2}(y)$, and the procedure can be iterated 
(in principle) to obtain all $A_{r,p}(y)$ coefficients in (2.13a),
while the $B_{r,p}(y)$ are obtained after imposing (2.14) and (2.15)
as we said before. The details of the construction are indeed a bit
involved, and hence it is worth showing what can be done with the
integral operator $T^{-2}$ given in (2.19). On the one hand, Eqs. (2.9a)
and (2.9b) lead to
$$ \eqalignno{
(T^{-2}f)(x)&=\int_{-\infty}^{x}dy \Bigr[A_{1,2}(y)+xA_{2,2}(y)
\Bigr]f(y) \cr
&+\int_{x}^{\infty}dy \Bigr[y+A_{1,2}(y)+x(A_{2,2}(y)-1)\Bigr]f(y).
&(2.20)\cr}
$$
On the other hand, by virtue of (2.19), $(T^{-2}f)(x)$ is also 
given by
$$
(T^{-2}f)(x)=\int_{-\infty}^{x}dy {1\over 4}h(y)
+\int_{x}^{\infty}dy \left(-{1\over 4}h(y)\right)
\eqno (2.21)
$$
where
$$
h(y) \equiv \int_{-\infty}^{y}f(z)dz-\int_{y}^{\infty}f(z)dz.
\eqno (2.22)
$$
Direct comparison of the representations (2.20) and (2.21) of
$(T^{-2}f)(x)$ yields therefore the equations
$$
\Bigr[A_{1,2}(y)+xA_{2,2}(y)\Bigr]f(y)={1\over 4}h(y)
\eqno (2.23)
$$
$$
\Bigr[y+A_{1,2}(y)+x(A_{2,2}(y)-1)\Bigr]f(y)=-{1\over 4}h(y).
\eqno (2.24)
$$
The addition of (2.23) and (2.24) leads to
$$
\Bigr[2A_{1,2}(y)+2xA_{2,2}(y)+y-x \Bigr]f(y)=0
\eqno (2.25)
$$
which is satisfied for all $f(y)$ if and only if
$$
2A_{1,2}(y)+y=0
\eqno (2.26)
$$
$$
(2A_{2,2}(y)-1)x=0.
\eqno (2.27)
$$
Such a system is solved by
$$
A_{1,2}(y)=-{y\over 2}
\eqno (2.28)
$$
$$
A_{2,2}(y)={1\over 2}
\eqno (2.29)
$$
which provides the desired explicit formula for the Green function
$G_{2}(x,y)$, upon insertion into (2.9a) and (2.9b).
\vskip 0.3cm
\leftline {\bf 3. Symbol and ellipticity}
\vskip 0.3cm
\noindent
Recall now that, if $L$ is a pseudo-differential operator defined
by a kernel $K$, this is related to the symbol $p(x,\xi)$ by
the equation [3]
$$
K(x,y)=(2\pi)^{-n} \int_{{\bf R}^{n}} 
{\rm e}^{i(x-y)\cdot \xi}p(x,\xi)d\xi.
\eqno (3.1)
$$
This equation can be inverted to give a very useful formula for
the symbol, i.e. (cf Eq. (2.1.36) in [3])
$$
p(x,\xi)=\int_{{\bf R}^{n}} {\rm e}^{-iz \cdot \xi}K(x,x-z)dz.
\eqno (3.2)
$$
Equation (3.2) is a key formula for our investigation, because the
ellipticity of $L$ is defined in terms of its symbol, as we know
from the introduction, following [1].

In our problem, which involves $x \in {\bf R}$, the integral (3.2)
reduces to
$$
p(x,\xi)=\int_{-\infty}^{\infty}{\rm e}^{-iz \xi}K(x,x-z)dz
\eqno (3.3)
$$
where the kernel $K(x,y)$ is expressed by (2.11)--(2.13), and we
have to check that the inequality (1.7) is satisfied for
$|\xi| \geq C_{0}$ to obtain ellipticity. Indeed, the symbol
(3.3) turns out to be
$$ \eqalignno{
p(x,\xi)&=-\int_{-\infty}^{\infty}{\rm e}^{-iz \xi}\delta'(z)dz \cr
&+\int_{-\infty}^{\infty}{\rm e}^{-iz \xi} \sum_{p=1}^{\infty}
u_{p}(x)\sum_{r=1}^{p}C_{r,p}(x-z)x^{r-1}dz
&(3.4)\cr}
$$
and hence obeys the inequality
$$ \eqalignno{
|p(x,\xi)| & \geq \left | \int_{-\infty}^{\infty}{\rm e}^{-i z \xi}
\delta'(z)dz \right | \cr
&- \left | \int_{-\infty}^{\infty} 
{\rm e}^{-iz \xi} \sum_{p=1}^{\infty}u_{p}(x)
\sum_{r=1}^{p}C_{r,p}(x-z)x^{r-1} dz \right | .
&(3.5)\cr}
$$
Of course, the first integral on the right-hand side of (3.5) becomes
meaningful within the framework of Fourier transform of 
distributions [13]. In the simplest possible terms, one has 
actually to consider the parameter-dependent integral 
(here $a >0$)
$$ \eqalignno{
I_{1,a}(\xi) & \equiv \int_{-\infty}^{\infty}{\rm e}^{-a z^{2}}
{\rm e}^{-iz \xi} \delta'(z) dz \cr
&=\int_{-\infty}^{\infty}\delta(z)(-2az-i \xi)
{\rm e}^{-a z^{2}-iz \xi}dz.
&(3.6)\cr}
$$
By virtue of the property defining the Dirac delta functional,
according to which [12] 
$$
(\delta,f)=f(0)
\eqno (3.7)
$$
the integral (3.6) equals $-i \xi$, and hence the first term on the
right-hand side of (3.5) equals $| \xi |$. Now we distinguish
two cases, depending on whether
$$
f(x,\xi) \equiv |\xi| -\left | \int_{-\infty}^{\infty}
{\rm e}^{-iz \xi} \sum_{p=1}^{\infty}u_{p}(x) \sum_{r=1}^{p}
C_{r,p}(x-z) x^{r-1} dz \right | 
\eqno (3.8)
$$
is positive or negative. If 
$$
f(x,\xi) > 0
\eqno (3.9)
$$
holds, the majorization (1.7) for the ellipticity of 
the restricted Lax operator
is satisfied provided that, for $| \xi | \geq C_{0}$,
$$
\left | \int_{-\infty}^{\infty} {\rm e}^{-i z \xi} \sum_{p=1}^{\infty}
u_{p}(x) \sum_{r=1}^{p}C_{r,p}(x-z)x^{r-1} dz \right |
\leq | \xi |  -C_{1}^{-1}(1+|\xi|)^{d}.
\eqno (3.10)
$$
In contrast, if
$$
f(x,\xi) < 0
\eqno (3.11)
$$
holds, the restricted Lax operator is elliptic provided that
$$ \eqalignno{
\; & \left | \int_{-\infty}^{\infty} {\rm e}^{-iz \xi} \sum_{p=1}^{\infty}
u_{p}(x) \sum_{r=1}^{p}C_{r,p}(x-z)x^{r-1} dz \right | 
\geq | \xi | -C_{1}^{-1}(1+|\xi|)^{d} \cr
& \geq C_{0}-C_{1}^{-1}(1+|\xi|)^{d}
&(3.12)\cr}
$$
for $| \xi | \geq C_{0}$.
If the order $d$ of the Lax operator is positive, we can further
write that, for $| \xi | \geq C_{0}$, the majorization (3.10)
becomes
$$
\left | \int_{-\infty}^{\infty} {\rm e}^{-iz \xi} \sum_{p=1}^{\infty}
u_{p}(x) \sum_{r=1}^{p}C_{r,p}(x-z)x^{r-1} dz \right |
\leq | \xi | -C_{1}^{-1}C_{0}^{d}.
\eqno (3.13)
$$

If we are interested in sufficient conditions we can point out that,
since the inequality (3.5) is always satisfied, whereas (1.7) 
only holds when $L$ is elliptic, the sufficient condition for 
ellipticity of the restricted Lax operator is expressed by
$$
C_{1}^{-1}(1+|\xi|)^{d} \leq |\xi| -\left | \int_{-\infty}^{\infty}
{\rm e}^{-iz \xi} \sum_{p=1}^{\infty}u_{p}(x)\sum_{r=1}^{p}
C_{r,p}(x-z)x^{r-1} dz \right | 
\eqno (3.14)
$$
for $|\xi| \geq C_{0}$. This leads in turn to the inequality (3.10).

To sum up, if the function $f: T^{*}({\bf R}) \rightarrow
{\bf R}$ defined by (3.8) has no zeros (which is already a
non-trivial requirement on the potentials $u_{p}$), the restricted 
Lax operator is elliptic provided that either (3.10) or (3.12) is
satisfied. The majorization (3.10) is further simplified in the
form (3.13) in case of positive order of the Lax operator.
A sufficient condition for ellipticity is given instead by
(3.14), which coincides with (3.10). In particular, when 
$|\xi|=C_{0}$ and the equality sign is chosen in (3.14), the 
order $d$ of the restricted Lax operator can be evaluated by the formula
$$
d={\log (C_{1}(C_{0}-I_{\xi})) \over \log(1+C_{0})}
\eqno (3.15)
$$
where
$$
I_{\xi} \equiv {\rm sup}_{x \in {\bf R}} \left | \int_{-\infty}^{\infty}
{\rm e}^{-iz \xi} \sum_{p=1}^{\infty} u_{p}(x)\sum_{r=1}^{p}
C_{r,p}(x-z) x^{r-1} dz \right | 
\eqno (3.16)
$$
bearing in mind that the modulus of $\xi$ equals the 
constant $C_{0}$.
\vskip 0.3cm
\leftline {\bf 4. Strong ellipticity}
\vskip 0.3cm
\noindent
In a thorough analysis of the ellipticity properties, strong
ellipticity should also be studied. For this purpose, following [3],
we assume that the symbol of the restricted Lax operator is 
{\it polyhomogeneous}, in that it admits an asymptotic expansion
of the form
$$
p(x,\xi) \sim \sum_{l=0}^{\infty}p_{d-l}(x,\xi)
\eqno (4.1)
$$
where each term $p_{d-l}$ has the homogeneity property
$$
p_{d-l}(x,\gamma \xi)=\gamma^{d-l}p_{d-l}(x,\xi)
\eqno (4.2)
$$
for $t \geq 1$ and $|\xi| \geq 1$. The {\it principal symbol}
$p^{0}$ of the Lax operator is then, by definition,
$$
p^{0}(x,\xi) \equiv p_{d}(x,\xi).
\eqno (4.3)
$$
{\it Strong ellipticity} is formulated in terms of the principal
symbol, because it requires that
$$
{\rm Re} \; p^{0}(x,\xi)={1\over 2}\Bigr[p^{0}(x,\xi)
+p^{0}(x,\xi)^{*}\Bigr] \geq c(x) |\xi|^{d}
\eqno (4.4)
$$
where $x \in {\bf R}, c(x) > 0$ and $|\xi| \geq 1$. In other words,
given a positive function $c$, the product $c(x) |\xi|^{d}$ should 
be always majorized by the real part of the principal symbol of the
restricted Lax operator. Indeed, the symbol (3.4) is such that
$$ \eqalignno{
\; & p(x,\gamma \xi)=-i\gamma \xi \cr 
& +\gamma^{-1} \int_{-\infty}^{\infty}
{\rm e}^{-iz\xi} \sum_{p=1}^{\infty}u_{p}(x)
\sum_{r=1}^{p}C_{r,p}\left(x-{z\over \gamma}\right)
x^{r-1}dz.
&(4.5)\cr}
$$
By virtue of (4.1), (4.2) and (4.5) we find that
$$ \eqalignno{
\; & -i\gamma \xi +\gamma^{-1} \int_{-\infty}^{\infty}
{\rm e}^{-iz\xi} \sum_{p=1}^{\infty}u_{p}(x)
\sum_{r=1}^{p}C_{r,p}\left(x-{z\over \gamma}\right)
x^{r-1}dz \cr
& \sim \sum_{l=0}^{\infty}\gamma^{d-l}p_{d-l}(x,\xi).
&(4.6)\cr}
$$
Moreover, the term on the right-hand side of (4.6) with $l=0$
should be the one occurring in the condition (4.4) for strong
ellipticity. A mathematical advantage of strong ellipticity lies in
the possibility of having a well defined functional trace of the
heat semigroup associated to the Lax operator [1, 3].
\vskip 0.3cm
\leftline {\bf 5. Behaviour of the ellipticity conditions
under KP flows}
\vskip 0.3cm
\noindent
To study the preservation (or violation) of the ellipticity
conditions under KP flows one has to analyze the following
problem: suppose that the conditions (3.10) or (3.12) are satisfied
for $t=0$. Are they still valid for all or some $t>0?$

This means that we consider again the potentials $u_{k}$ as in
Eq. (1.11), i.e. as functions of $x$ and $t$, where $t$ is a
concise notation for infinitely many time parameters 
$(t_{1},t_{2},...)$. It should be stressed that we consider
only one spatial variable and infinitely many time parameters,
since otherwise it would be problematic, at least for the authors,
to generalize formulae like (3.2) aimed at obtaining the 
symbol from the kernel of the operator.
For our purposes it is convenient to use
formulae generating such potentials by means of a single function.
This is made possible by the $\tau$-function (see appendix),
because one finds [10]
$$
u_{1}(x;t)={\partial^{2}\over \partial x^{2}} \log \tau
\eqno (5.1)
$$
$$
u_{2}(x;t)={1\over 2}\left(-{\partial^{3}\over \partial x^{3}}
+{\partial^{2}\over \partial x \partial t_{2}}\right)
\log \tau
\eqno (5.2)
$$
$$
u_{3}(x;t)={1\over 6}\left({\partial^{4}\over \partial x^{4}}
-3{\partial^{3}\over \partial x^{2} \partial t_{2}}
+2{\partial^{2}\over \partial x \partial t_{3}}\right)
\log \tau -u_{1}^{2}
\eqno (5.3)
$$
and infinitely many other equations of the general form
$$
u_{k}(x;t)=F_{k}(\log \tau)
\eqno (5.4)
$$
where $F_{k}$ is, in general, a non-linear function of $\log \tau$.
Equations (5.4), with $k$ ranging from $1$ through $\infty$, should
be inserted into the ellipticity conditions (3.10) and (3.14),
substituting therein $u_{k}$ with $F_{k}(\log \tau)$ for all $k$.
The resulting majorizations involve non-linear functions of the
logarithm of the $\tau$-function.

Further progress can be made by considering a `truncated' Lax
operator, e.g.
$$
{\widetilde L} \equiv T+u_{1}(x;t)T^{-1}+u_{2}(x;t)T^{-2}.
\eqno (5.5)
$$
This should not seem an arbitrary simplification, because the
Lax operator is obtained from the $W$ operator of the appendix
as [10]
$$
L \equiv W \; T \; W^{-1}.
\eqno (5.6)
$$
Now both $W$ and its `truncated version' [10]
$$
W_{m} \equiv 1+\sum_{k=1}^{m}w_{k}T^{-k}
\eqno (5.7)
$$
satisfy the Sato equation (A8), from which the generalized Lax
equation (1.16) is eventually obtained. Thus, operators like
$\widetilde L$ in (5.5) can be obtained from (5.6) if $W$ is
replaced by $W_{m}$ therein. In this case, on defining
$$ \eqalignno{
\; & F(x,z;t) \equiv \sum_{p=1}^{2}u_{p}(x;t)\sum_{r=1}^{p}
C_{r,p}(x-z)x^{r-1} \cr
&=u_{1}(x;t)C_{1,1}(x-z)
+u_{2}(x;t)\Bigr[C_{1,2}(x-z)+xC_{2,2}(x-z)\Bigr]
&(5.8)\cr}
$$
the integral on the right-hand side of the ellipticity condition (3.14)
reduces to $(J_{1}+J_{2}+J_{3})(x,\xi;t)$ where, bearing in mind that
(see (2.17), (2.28) and (2.29))
$$
C_{1,1}(x-z)=C_{2,2}(x-z)={1\over 2} \; {\rm if} \; z>0, 
\; -{1\over 2} \; {\rm if} \; z<0
\eqno (5.9)
$$
$$
C_{1,2}(x-z)=-{x\over 2}+{z\over 2} \; {\rm if} \; z>0,
\; {x\over 2}-{z\over 2} \; {\rm if} \; z<0
\eqno (5.10)
$$
one finds
$$ \eqalignno{
\; & J_{1}(x,\xi;t) \equiv \int_{-\infty}^{\infty}
{\rm e}^{-iz \xi}u_{1}(x;t)C_{1,1}(x-z)dz \cr
&=u_{1}(x;t)\lim_{b \to \infty} \left({2\over \xi}
{\rm e}^{-i b {\xi \over 2}} \sin {b \xi \over 2} \right)
&(5.11)\cr}
$$
$$ \eqalignno{
\; & J_{2}(x,\xi;t) \equiv \int_{-\infty}^{\infty}
{\rm e}^{-iz \xi}u_{2}(x;t)C_{1,2}(x-z)dz \cr
&=-xu_{2}(x;t){J_{1}(x,\xi;t)\over u_{1}(x;t)}
+u_{2}(x;t)i{d\over d\xi}{J_{1}(x,\xi;t)\over u_{1}(x;t)}
&(5.12)\cr}
$$
$$ \eqalignno{
\; & J_{3}(x,\xi;t) \equiv \int_{-\infty}^{\infty}
{\rm e}^{-iz \xi}u_{2}(x;t)xC_{2,2}(x-z)dz \cr
&=xu_{2}(x;t){J_{1}(x,\xi;t)\over u_{1}(x;t)}.
&(5.13)\cr}
$$
In these equations, the infinite upper limit of integration can be
recovered by taking the limit as $b \rightarrow \infty$ of integrals
from $0$ to $b$ (the lower limit being amenable to $0$ by virtue
of (5.9) and (5.10). However, divergences remain, and hence we are
only able to obtain well defined formulae by integrating up to
finite values of $b$. The cancellation of terms involving
$xu_{2}(x;t)$ is thus found to occur on performing the sum, and
our integral reads eventually
$$
(J_{1}+J_{2}+J_{3})_{b}(x,\xi;t)=u_{1}(x;t)F_{b}(\xi)
+u_{2}(x;t)i{d\over d\xi}F_{b}(\xi)
\eqno (5.14)
$$
having defined
$$
F_{b}(\xi) \equiv {2\over \xi}
{\rm e}^{-ib{\xi \over 2}}\sin {b \xi \over 2}.
\eqno (5.15)
$$
Now we point out that
$$ \eqalignno{
\; & \left | u_{1}(x;t)F_{b}(\xi)+u_{2}(x;t)iF_{b}'(\xi)
\right | \leq \left | u_{1}(x;t)F_{b}(\xi) \right|
+ \left | u_{2}(x;t) i F_{b}'(\xi) \right | \cr
& \leq {2\over \xi} \left | u_{1}(x;t) \right |
+ \left({b\over \xi}+{2\over \xi^{2}}\right) 
\left | u_{2}(x;t) \right| .
&(5.16)\cr}
$$
Thus, a sufficient condition for the
validity of the majorization (3.14) is expressed by
$$
| \xi | \geq C_{1}^{-1} (1+|\xi|)^{d}+{2\over \xi}|u_{1}(x;t)|
+ \left({b \over \xi}+{2\over \xi^{2}}\right) |u_{2}(x;t)|.
\eqno (5.17)
$$
It should be stressed that divergent integrals occur already in the
`time-independent' ellipticity condition (3.14). When the time
parameters are introduced, it may be easier or harder to fulfill
ellipticity, depending on the behaviour of $u_{1},u_{2},...$
(which, in turn, all depend on the $\tau$-function).
\vskip 0.3cm
\leftline {\bf 6. Concluding remarks}
\vskip 0.3cm
\noindent
Our paper has been motivated by the need to obtain a deeper
understanding of the basic structures of modern non-linear physics.
It is indeed well known that the Sato equation (A8) generates the
generalized Lax equation (1.16), the Zakharov--Shabat equation 
(1.17) and the inverse spectral transform scheme [10]. Moreover,
an infinite number of nonlinear evolution equations (i.e. the
KP hierarchy), of which the KP equation is the simplest nontrivial
one, share solutions, and the $\tau$-function makes it possible
to express all such solutions. 

The work in the present paper is the first step towards a rigorous
investigation of the ellipticity properties of the Lax
pseudo-differential operator. We have found that, to achieve
ellipticity, including its strong form, 
the various potentials $u_{k}(x)$ are no
longer arbitrary, but should be chosen in such a way that the
following conditions hold:
\vskip 0.3cm
\noindent
(i) The function $f: T^{*}{\bf R} \rightarrow {\bf R}$ defined
in (3.8) has no zeros.
\vskip 0.3cm
\noindent
(ii) The majorization (3.14) holds (some care is actually necessary
to deal with the integrals on the right-hand side of (3.14), as
is clear from the analysis performed in section 5).
\vskip 0.3cm
\noindent
(iii) The asymptotic expansion (4.6) can be obtained. Note, however,
that violation of (4.2) (i.e. lack of homogeneity) for
$|\xi| < 1$ can cause logarithmic terms in the asymptotic expansion
of the kernel defined by (2.11)--(2.13) (cf [14]).
\vskip 0.3cm

Moreover, on allowing for the time evolution of the potentials in
the Lax operator, now viewed as functions of $x$ and of infinitely
many time variables, we have obtained an explicit ellipticity
condition in terms of the $\tau$-function, when attention is restricted
to the `truncated' Lax operator (5.5). It now remains to be seen
how to deal with the infinite sum over $p$ in the ellipticity
condition (3.14) when the potentials $u_{p}(x;t)$ appropriate for the
`full' Lax operator (1.11) are instead considered. The form of $u_{1}$
and $u_{2}$ remains the one given in (5.1) and (5.2), but the occurrence
of an infinite number of such potentials makes it hard to re-express
the time-dependent form of the majorization (3.14).
The work in [15] has indeed obtained a very useful formula for the
$\tau$-function, but this remains of little help when the infinite
sum over all potentials is performed.

A further interesting issue is the investigation of ellipticity
for the formulation of KP hierarchy considered in [16], where 
the main new technique, when compared to the traditional approach
to the generalized Lax equation, consists of replacing the Lax
operator by an $n^{\rm th}$-order formal pseudo-differential
operator
$$
L_{n} \equiv T^{n}+\sum_{j=-\infty}^{n-2}q_{j}T^{j} \; \;
n \geq 2.
\eqno (6.1)
$$
The authors of [16] have been able to factorize $L_{n}$ into $n-1$
first-order formal differential operators $A_{k}, 1 \leq k \leq n-1$,
and one first-order formal pseudo-differential operator 
${\widetilde A}_{n}$, i.e.
$$
L_{n}={\widetilde A}_{n}A_{n-1}...A_{2}A_{1}
\eqno (6.2)
$$
where
$$
A_{k} \equiv T+\eta_{k,x} \; \; 1 \leq k \leq n
\eqno (6.3)
$$
$$
\sum_{k=1}^{n}\eta_{k,x}=0
\eqno (6.4)
$$
$$
{\widetilde A}_{n} \equiv A_{n}+\sum_{j=-\infty}^{-1}
b_{n,j}T^{j}.
\eqno (6.5)
$$

The results and unsolved problems described so far seem to show
that new exciting developments might be obtained from the effort
of combining some key techniques of non-linear physics and the 
tools of linear and pseudo-differential operator theory.
In particular, the mathematical requirement of ellipticity in the
various forms considered in sections 3--5 restricts the potentials
$u_{k}$ in a form not previously considered in the literature to
our knowledge, which might be used to select the realizations of
the Lax operator one is interested in.
\vskip 0.3cm
\leftline {\bf Acknowledgment}
\vskip 0.3cm
\noindent
This work has been partially supported by PRIN97 `Sintesi'.
Correspondence with Gerd Grubb has been helpful for the authors,
as well as constant encouragement from Giuseppe Marmo.
\vskip 0.3cm
\leftline {\bf Appendix}
\vskip 0.3cm
\noindent
Since the general reader is not necessarily familiar with the theory
of $\tau$-functions, we summarize the main properties hereafter.
Following [10], we study for the operator $W_{m}$ defined in (5.7)
the ordinary differential equation
$$
W_{m}\partial^{m}f(x)=(\partial^{m}+w_{1}(x)\partial^{m-1}+...
+w_{m}(x))f(x)=0
\eqno (A1)
$$
which has $m$ linearly independent solutions $f^{(1)}(x),...,f^{(m)}(x)$.
On writing equation (A1) $m$ times with $f=f^{(1)}(x),...,f=f^{(m)}(x)$,
one finds a linear system which can be solved for $w_{j}(x)$, for all
$j=1,...,m$, and hence $W_{m}$ is found from the definition (5.7).

When the $w_{j}$ are taken to depend also on infinitely many time
variables $(t_{1},...,t_{p},...)$, the operator $W_{m}(x;t)$ is
found to be 
$$
W_{m}(x;t)={{\rm det}A \over \tau(x;t)}
\eqno (A2)
$$
where, given the functions $h_{0}^{(j)}(x;t)$ and $h_{n}^{(j)}(x;t)$ 
such that
$$
\left({\partial \over \partial t_{n}}-{\partial^{n}\over \partial x^{n}}
\right)h_{0}^{(j)}(x;t)=0 \; \; \; n=1,2,...
\eqno (A3)
$$
$$
h_{0}^{(j)}(x;0)=f^{(j)}(x)
\eqno (A4)
$$
$$
h_{n}^{(j)}(x;t)={\partial \over \partial t_{n}}h_{0}^{(j)}(x;t)
={\partial^{n}\over \partial x^{n}}h_{0}^{(j)}(x;t)
\eqno (A5)
$$
one has (see the definition (1.10))
$$
A \equiv \pmatrix{h_{0}^{(1)} & ... & h_{0}^{(m)} & T^{-m} \cr
... & ... & ... & ... \cr
h_{m-1}^{(1)} & ... & h_{m-1}^{(m)} & T^{-1} \cr
h_{m}^{(1)} & ... & h_{m}^{(m)} & 1 \cr}
\eqno (A6)
$$
and
$$
\tau(x;t) \equiv {\rm det} 
\pmatrix{h_{0}^{(1)} & ... & h_{0}^{(m)}\cr
... & ... & ... \cr
h_{m-1}^{(1)} & ... & h_{m-1}^{(m)} \cr}.
\eqno (A7)
$$
The function $\tau$ is the $\tau$-function used in Eqs. (5.1)--(5.4),
and the time evolution of the operator $W_{m}(x;t)$ is determined 
by the Sato equation
$$
{\partial W_{m}\over \partial t_{n}}=B_{n}W_{m}-W_{m}T^{n}
\eqno (A8)
$$
where the operators $B_{n}$ are the same occurring in the generalized
Lax equation (1.16).
\vskip 0.3cm
\leftline {\bf References}
\vskip 0.3cm
\noindent
\item {[1]}
Gilkey P B 1995 {\it Invariance Theory, the Heat Equation and the
Atiyah--Singer Index Theorem} (Boca Raton, FL: Chemical Rubber 
Company)
\item {[2]}
Seeley R T 1969 {\it Topics in pseudo-differential operators,
C.I.M.E.}, in Conf. on Pseudo-Differential Operators, ed. L.
Nirenberg (Roma: Edizioni Cremonese)
\item {[3]}
Grubb G 1996 {\it Functional Calculus of Pseudodifferential
Boundary Problems (Progress of Mathematics 65)}
(Boston: Birkh\"{a}user)
\item {[4]}
Esposito G 1999 {\it Class. Quantum Grav.} {\bf 16} 1113
\item {[5]}
Esposito G 1999 {\it Class. Quantum Grav.} {\bf 16} 3999
\item {[6]}
Esposito G and Stornaiolo C 2000 {\it Int. J. Mod. Phys.}
{\bf A15} 449
\item {[7]}
Ablowitz M J and Clarkson P A 1991 {\it Solitons, Nonlinear
Evolution Equations and Inverse Scattering} (Cambridge:
Cambridge University Press)
\item {[8]}
Zakharov V E 1991 {\it What is Integrability?} 
(Berlin: Springer-Berlag)
\item {[9]}
Date E, Kashiwara M, Jimbo M and Miwa T 1983 Transformation groups
for soliton equations, in Proc. of RIMS Symposium on Nonlinear
Integrable Systems, Classical Theory and Quantum Theory, eds
M Jimbo and T Miwa (Singapore: World Scientific)
\item {[10]}
Ohta Y, Satsuma J, Takahashi D and Tokihiro T 1988
{\it Prog. Theor. Phys. Suppl.} {\bf 94} 210
\item {[11]}
Lanczos C 1961 {\it Linear Differential Operators}
(London: Van Nostrand)
\item {[12]}
Roos B W 1969 {\it Analytic Functions and Distributions in
Physics and Engineering} (New York: John Wiley)
\item {[13]}
Kolmogorov A N and Fomin S V 1980 {\it Elements of the Theory
of Functions and Functional Analysis} (Moscow: Mir)
\item {[14]}
Grubb G and Seeley R T 1995 {\it Inv. Math.} {\bf 121} 481
\item {[15]}
Chau L L, Shaw J C and Yen H C 1992 {\it Commun. Math. Phys.}
{\bf 149} 263
\item {[16]}
Gesztesy F and Unterkofler K 1995 {\it Diff. Int. Eqs.}
{\bf 8} 797

\bye